\documentstyle[twocolumn,prl,aps,psfig]{revtex}
\begin{document}
\twocolumn[\hsize\textwidth\columnwidth\hsize\csname@twocolumnfalse%
\endcsname
\title{Fano Lineshapes Revisited: Symmetric Photoionization Peaks
from Pure Continuum Excitation}
\author{U. Eichmann$^{1,2}$, T.F. Gallagher$^{1,3}$, and R.M. Konik$^{3}$}
\address{
$^1$Max-Born-Institute for Nonlinear Optics and Short
Pulse Spectroscopy, D-12489 Berlin, Germany}
\address{
$^2$Institut f\"ur Atomare Physik und Fachdidaktik,
Technical University Berlin, D-10623 Berlin, Germany}
\address{$^3$
Department of Physics, University of Virginia, Charlottesville, USA}

\date{February 4, 2003}
\maketitle
\begin{abstract}
In a photoionization spectrum in which there is no excitation of the 
discrete states, but only the underlying continuum, we have observed 
resonances which appear as symmetric peaks, not the commonly expected 
window resonances.  Furthermore, since the excitation to the unperturbed 
continuum vanishes, the cross section expected from Fano's configuration 
interaction theory is identically zero.  This shortcoming is removed by 
the explicit introduction of the phase shifted continuum, which demonstrates 
that the shape of a resonance, by itself, provides no information about 
the relative excitation amplitudes to the discrete state and the continuum.
\end{abstract}
\pacs{PACS numbers: ???}  ]  
\bigskip
\vskip .4in

\newcommand{\del}{\partial}

Quantum interference occurs whenever there exist two coherent paths 
from an initial state to a final state.  Particularly fascinating 
is the case in which one of the two paths is via a resonance, for in 
this case the presence of the resonance is manifested in a wide variety of 
lineshapes. In the case of optical absorption they are often termed Fano 
lineshapes\cite{fano}.
One of the earliest examples occurred in the absorption 
spectrum of Ar, Kr, and Xe\cite{beutler}.
Above the first ionization limit the 
rare gas atoms can be photoionized either directly or via the doubly 
excited states, which are coupled to the ionization continuum.  The 
absorption cross section due to the doubly excited states does not simply 
add to the continuum photoionization cross section, as might be naively 
expected for a Breit-Wigner resonance\cite{blatt}.
Rather, the amplitudes for 
excitation of the doubly excited state and the continuum must be added, 
often leading to asymmetric resonances.  Such asymmetric resonances are 
ubiquitous in the photoionization of atoms and molecules\cite{ares},
and their 
existence prompted Fano to develop his seminal theory of configuration 
interaction between a discrete state and a continuum.

As it becomes possible to preserve quantum mechanical coherence in 
more complex systems, it is likely that Fano's theory will find 
increasingly wide application.  For example, photoabsorption in quantum 
well systems exhibits interference which is essentially identical to 
that observed in atomic photoionization\cite{qw1,qw2}.  Somewhat different 
manifestations occur in the conductance through magnetic impurity 
atoms\cite{kondo} and single electron transistors\cite{fanodot}.
Extensions of 
Fano's theory have been worked out for these problems\cite{fanoexact} 
and for its application to chaotic systems\cite{fanochaos}.

Here we report an experiment which reveals a shortcoming of the 
straightforward application of Fano's theory.  In particular we 
describe a photoionization experiment in which the excitation amplitudes 
to a series of discrete states vanish, yet we see symmetric peaks at 
their locations, not the commonly expected window resonances, or dips, 
in the photoionization cross section.  More problematic, the excitation 
amplitude to the unperturbed continuum also vanishes, leading the theory 
to predict no excitation at all.  In fact, the theory is not 
completely correct for long range coulomb potentials and thus fails 
to describe photoionization.  In the sections which follow we describe 
our experiment, review Fano's theory, point out the source of the 
problem, and suggest the correct form for the photoionization cross section.

\begin{figure}[tbh]
\vskip -.7in
\centerline{\psfig{figure=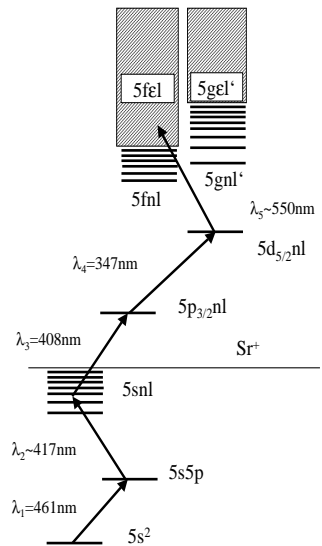,height=3.9in,width=2.4in}}
\caption{Excitation scheme of the experiment.  
The excitation of the $Sr5d17l$
state is done with four fixed frequency lasers and Stark switching.  
The frequency of the final fifth laser is swept through the energy from the
$Sr^+5f$ to $5g$ limits.  
As shown, there is no excitation amplitude to the
$5gnl'$ states, only to the $5f\epsilon l$ continuum.}
\end{figure}

In the experiment Sr atoms in a beam are excited to the doubly excited
$5d17l$ state with $l=12$
using four pulsed lasers and a Stark switching technique, as shown 
in the energy level diagram of Fig. 1 and described in detail 
elsewhere\cite{expdetail1}.
Sr atoms in the $5d17l$
state are then exposed to a fifth, ~550 nm, laser pulse which 
excites them to the energy range between the $Sr^+5f$ and $5g$
ionization limits.
This excitation could imaginably produce
either directly a $Sr^+5f$ ion together with a free electron
or a $5gnl'$ atom.  This latter state 
autoionizes quickly (in roughly 1 ns), thus 
producing a free electron and so would again leave
the ion predominantly in the excited $Sr^+5f$ state. The same 550nm laser
pulse then ionizes the $Sr^+5f$ ion to produce $Sr^{++}$\cite{expdetail2}.
The production of
$Sr^{++}$ is proportional to the excitation by the first 550 nm
photon. 
Approximately $100ns$
after the laser pulse we apply a 1 kV/cm electric field pulse which 
drives the $Sr^{++}$ ions to a dual microchannel plate detector.  
The detector signal is recorded with a gated integrator as the wavelength 
of the 550 nm laser is slowly scanned over many shots of the lasers.  
The observed photoionization spectrum is shown in Fig. 2.  In it a 
clear series of symmetric, apparently Lorentzian peaks, at the locations 
of the $5gnl'$ states ($l'=11,13$),
is quite evident, and there is no photoionization between the peaks.  
At first glance it seems obvious that we are only exciting 
the $5gnl'$ states and not exciting
the $5f\epsilon l$
continuum at all.  
However, after more careful consideration it becomes apparent that 
quite the opposite is true.  The initial $5d17l$
state is well represented by an independent particle picture, i.e., 
a $Sr^{+}5d$ ion with a hydrogenic $nl$
electron bound to it, and the wavefunction is the product of these 
two wavefunctions.  There is evidently no electric dipole coupling 
from the $5d17l$ state to the $5gnl'$ state.
In contrast, the dipole coupling from the $5d17l$ state to the
$5f\epsilon l$ 
continuum is allowed.  In particular the $Sr^{+}$ ion makes the 
$5d-5f$ transition, and the $nl$
spectator electron is shaken off to the
$\epsilon l$ continuum, resulting in the
$5f\epsilon l$ final state\cite{gallagher1,gallagher2,gallagher3}.
However, shake off to the unperturbed hydrogenic 
$5f\epsilon l$ continuum, $\psi_E$
is everywhere forbidden, and we only observe the excitation to the
$5f\epsilon l$ continuum where it is phase shifted by its 
interaction with the $5gnl'$ states.

Fano's theory describes the excitation from an initial state $i$ 
to a final state $f$, which consists of a discrete state 
$\phi$ at energy $E_\phi$ and the degenerate continuum 
$\psi_E$, which we assume to be energy normalized.  
(We follow the notation of ref. 1.)  It is most often the case that 
the coupling from the discrete state to the continuum, $V_E$, 
is energy independent, and we here consider this case. 
This coupling broadens the discrete state so that it has a width (FWHM), 
$\Gamma = 2\pi|V_E|^2$,
and the natural energy scale for the problem is the reduced energy, 
$\epsilon = 2(E-E_\phi )/\Gamma$.
In addition to broadening the discrete state, $\phi$,
the coupling $V_E$ also produces a phase shift $\Delta$
in the radial phase of the continuum wave function, 
and as we pass from far below to far above the discrete state at
$E_\phi$ there is a phase shift of $\pi$.
In particular, $\Delta$ is given by 
$\Delta = \cot^{-1}(\epsilon)$,
so that far below, at, and far above the resonance at 
$E_\phi$, $\Delta = 0,\pi/2$, and $\pi$, respectively. 
Well removed from the resonance the continuum wavefunction is described 
by its unperturbed solution $\psi_E \sim \sin(kr+\phi_{bg})$
where k is the continuum electron's wave number and 
$\phi_{bg}$ is a background phase.  At the resonance, $E_\phi$,
it is described by its phase shifted solution, $\Lambda_E$
with asymptotic form $\sim \cos(kr+\phi_{bg})$,
and in general by $\psi_E\cos (\Delta ) + \Lambda_E \sin (\Delta )$
with asymptotic form $~\sin (kr+\phi_{bg}+\Delta )$.\cite{fano,friedrich}
The resulting continuum wavefunction has the same asymptotic 
amplitude across the resonance.  We note that these forms of the 
continuum wavefunction, given in ref. 1, are appropriate for 
short range potentials.

\begin{figure}[tbh]
\vskip -.6in
\centerline{\psfig{figure=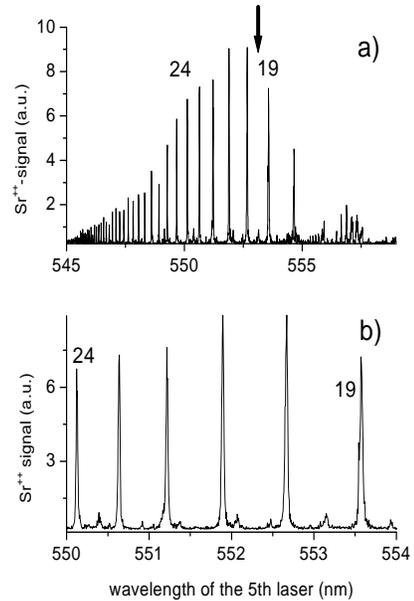,height=3.9in,width=2.4in}}
\caption{
Photoionization spectrum observed by scanning the fifth 
laser (a) from slightly below the $5f$
limit, shown by the arrow, to the vicinity of the $5g$ limit at 
about $544 nm$.  The resonances corresponding to 
$5g19l'$ and $5g24l'$
states have the numbers 19 and 24, respectively, above them. 
(b) expanded view of the $5g19l'-5g24l'$ resonances.}
\end{figure}

The photoexcitation cross section is composed of the excitation amplitudes 
to the discrete state $\phi$, the unperturbed continuum $\psi_E$, and 
the phase shifted continuum $\Lambda_E$.
In Fano's theory the phase shifted continuum 
is represented as a principal part integral over the unperturbed 
continuum, yielding the following expression,
\begin{eqnarray}\label{1}
\sigma &\propto& \bigg| 
{\langle\phi | \mu | i\rangle \sin \Delta \over  \pi V^*_E} 
+ {1 \over \pi V^*_E} 
{\bf P}\int dE' {V^*_{E'} \langle \psi_{E'}|\mu | i\rangle \over (E-E')}\sin\Delta\cr
&&\hskip 1.75cm  - \langle \psi_E|\mu | i\rangle \cos\Delta\bigg|^2 ,
\end{eqnarray}
where $\mu$ is the transition electric dipole moment,
$\langle \phi | \mu | i\rangle$ and $\langle \psi_E | \mu | i\rangle$
are the excitation matrix elements to the discrete state and the unperturbed 
continuum, and $\bf P$ denotes a principal part integral.

If the excitation amplitude to the continuum $\langle \psi_E|\mu|i\rangle$
is assumed to be energy independent, it appears that the principal part 
integral can be neglected, and doing so leads to the following common 
misinterpretation of the theory.  Namely, if there is no continuum 
excitation, there is a symmetric, approximately 
Lorentzian peak of width
$\Gamma$ centered at $E_\phi$.
On the other hand, if there is no excitation of the discrete state, 
$\langle\phi |\mu | i\rangle$,
there is a symmetric dip, or window resonance, 
in the photoionization cross section with vanishing excitation at $E_\phi$.
If both amplitudes are non zero the resulting interference term 
leads to the familiar asymmetric Fano lineshape.

In addition to being a source of confusion, writing the continuum $\Lambda_E$
as the principal part integral of Eq. (1) is incorrect for long 
range potentials which support bound states.  It thus does not correctly 
represent photoionization, as shown graphically by our experiment. 
However, it does represent $\Lambda_E$
correctly for short range potentials, 
as encountered in photodetachment\cite{ph_det}.  
There are two straightforward 
ways to remedy this shortcoming of Fano's theory.  
The first is 
to extend the principal part integral of Eq. (1) so that it includes 
not only the unperturbed continuum $\psi_E$
(here $5f\epsilon l$),
but the associated bound states as well (here $5fnl$)\cite{theory}.
This extension ultimately reflects the fact that the set of continuum states
(here $5f\epsilon l$) are not by themselves complete in terms of
the radial functions.
The second is to adopt a more physical approach and rewrite Eq. (1) 
using the phase shifted continuum explicitly, i.e.
\begin{eqnarray}\label{2}
\sigma \hskip -.5mm \propto \hskip -.5mm \bigg| 
{\langle\phi | \mu | i\rangle \sin \Delta \over  \pi V^*_E} 
\hskip -.5mm - \hskip -.5mm \langle \psi_E|\mu | i\rangle \cos\Delta 
 -\langle \Lambda_E|\mu | i\rangle \sin\Delta\bigg|^2 .
\end{eqnarray}
In this form it is apparent that discarding the principal part integral 
is equivalent to neglecting the excitation amplitude to the phase 
shifted continuum, $\Lambda_E$,
which is likely to be comparable to or greater than the excitation 
amplitude to the unperturbed continuum $\psi_E$.

The most physically appealing way of writing Eq. (2) is to assume a 
sinusoidal dependence of the continuum excitation amplitude on the 
phase $\Delta$ and replace
$\langle \psi_E|\mu | i\rangle \cos\Delta 
+\langle \Lambda_E|\mu | i\rangle \sin\Delta$ in Eq. (2) by
$\langle \psi_E|\mu | i\rangle_{\rm max}\cos (\Delta-\phi_i)$.
Here $\langle \psi_E|\mu | i\rangle_{\rm max}$
is the maximum transition amplitude from $i$ to the continuum as a 
function of the radial continuum phase, and $\phi_i$
is a measure of the radial phase difference between the initial state 
and the unperturbed continuum $\psi_E$.
$\langle \psi_E|\mu | i\rangle_{\rm max}$, assumed to be positive,
decreases slowly with energy.  
Far from the resonance, where $\Delta=0$ or $\pi$, the excitation amplitude 
to the continuum takes the value 
$\pm\langle \psi_E|\mu | i\rangle_{\rm max}\cos\phi_i$,
which is in general smaller in magnitude than 
$\langle \psi_E|\mu | i\rangle_{\rm max}$.
With this modification the cross section is given by 
\begin{eqnarray}\label{3}
\sigma \hskip -.5mm \propto \hskip -.5mm \bigg| 
{\langle\phi | \mu | i\rangle \sin \Delta \over  \pi V^*_E} 
 -\langle \psi_E|\mu | i\rangle_{\rm max} \cos(\Delta -\phi_i)\bigg|^2 .
\end{eqnarray}
With no excitation of the discrete state, i.e., 
$\langle\phi | \mu | i\rangle = 0$, 
and only continuum excitation, it is clear that any lineshape 
can be obtained using Eq. (3), and several are shown in Fig. 3 for 
different positive values of $\phi_i \leq \pi/2$.
For negative values the profiles are reflected through $\epsilon = 0$.
As shown by Fig. 3, $\phi_i = \pi/2$
leads to symmetric peaks, as seen in our spectrum of Fig. 2.

\begin{figure}[tbh]
\centerline{\psfig{figure=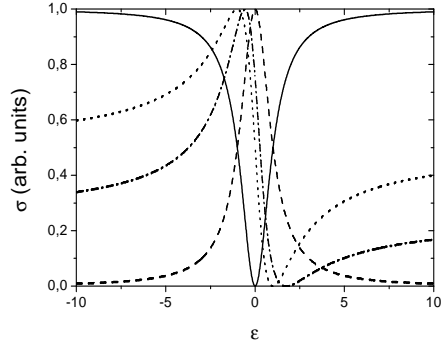,angle=-90,height=3.in,width=2.in}}
\vskip -.8in
\caption{
Relative cross sections for pure continuum excitation assuming the 
same value of $\langle \psi_E|\mu | i\rangle_{\rm max}$,
or equivalently, the maximum cross section, in all cases.  
Four values of the initial state continuum phase are shown; 
$\phi_i=0,\pi/4,\pi/3,$ and $\pi/2$ 
corresponding to $q = 0, -1, -1.732,$ and -$\infty$.  For $\phi < 0$
the profiles are reflected through $\epsilon = 0$.}
\end{figure}

Why the spectrum of Fig. 2 is a case in which $\phi_i=\pi/2$
is easily understood.  If we consider any one of the peaks of Fig. 2, 
the discrete state $\phi$ 
is the $5gnl'$
state and the unperturbed continuum $\psi_E$
is the $5f\epsilon l$ continuum.  As we have already stated, the dipole 
moment from the initial state, $i=5d17l$,
to the discrete state, $\phi=5gnl'$, vanishes. The dipole matrix element 
for excitation from the $5d17l$ 
state to the $5f\epsilon l$ continuum is given by
\begin{equation}\label{4}
\langle 5f\epsilon l|\mu | 5d 17l\rangle = \langle5f |\mu |5d\rangle
\langle \epsilon l | 17l\rangle,
\end{equation}
i.e., a product of the ionic dipole matrix element and an overlap integral 
for the outer electron.  In both the unperturbed $5d17l$
state and the unperturbed $5f\epsilon l$
continuum, the outer ($17l$ or $\epsilon l$)
electron states are hydrogenic and have quantum defects 
$\delta = 0$.
Consequently, the overlap integral $\langle 17l|\epsilon l\rangle$
vanishes due to the orthogonality of the $17l$ and $\epsilon l$
radial wavefunctions.  Evidently, with no interaction between the discrete 
state and the continuum there would be no continuum excitation.  
However, when the configuration interaction is taken into account the 
continuum excitation is allowed.  Specifically, as we pass the energies 
of the $5gnl'$ states the interaction of the $5f\epsilon l$
continuum with the $5gnl'$ states causes the radial continuum 
phase to go through a phase shift of $\pi$.
The change in the radial phase causes the overlap integral of Eq. (4) 
to depart from zero.  In particular, it oscillates sinusoidally with 
$\Delta$, having the form
\begin{equation}\label{5}
\langle \epsilon l|17l\rangle = \langle \epsilon l | 17l\rangle_{\rm max}
\sin\Delta ,
\end{equation}
where $\langle \epsilon l | 17l\rangle_{\rm max}$
decreases slowly with increasing energy.  Clearly the overlap integral 
reaches its maximum at $\Delta = \pi/2$, the location of the 
$5gnl'$ states.  With this observation we can use Eqs. (4) and (5) to 
write the second term of Eq. (3) for our spectrum of Fig. 2 as
\begin{eqnarray}\label{6}
\langle \psi_E| \mu | i\rangle_{\rm max}\cos (\Delta-\phi_i )
&=& \cr
&& \hskip -2cm 
\langle 5f | \mu | 5d\rangle \langle \epsilon l|17d\rangle_{\rm max}
\cos (\Delta - \pi/2 ),
\end{eqnarray}
i.e., the spectrum of Fig. 2 corresponds to the 
$\phi_i = \pi/2$ case shown in Fig. 3.  
In spite of the fact that the peaks of Fig. 2 appear to be from 
excitation of the discrete $5gnl'$
states, they are due only to the phase shifted $5f\epsilon l$
continuum.  Asymmetric lineshapes attributed to pure continuum excitation 
have been observed previously, but it was less clear in those cases that 
there was no excitation to the discrete state\cite{gallagher2,gallagher3}.

It is conventional to express the shape of the resonance as the ratio 
of the photoionization cross section to the cross section of the 
unperturbed continuum, i.e. as
\begin{equation}\label{7}
\bigg| {\langle f | \mu | i\rangle \over \langle \psi_E | \mu | i \rangle }
\bigg|^2 = {(q+\epsilon)^2 \over 1 + \epsilon^2},
\end{equation}
where we have introduced the Fano shape parameter q, defined 
as \cite{fano,friedrich}
\begin{equation}\label{8}
q = {\langle \phi | \mu | i \rangle /\pi V^*_E - 
\langle \psi_E|\mu | i\rangle_{\rm max}\sin\phi_i \over 
\langle \psi_E|\mu | i\rangle_{\rm max}\cos\phi_i }.
\end{equation}
It is evidently minus the ratio of the coefficients of the
$\sin\Delta$ and $\cos\Delta$ 
terms of Eqs. (1) or (2).  If we use the original Fano form of Eq. (1) 
and ignore the principal part integral, the second term in the numerator 
of Eq. (8) is missing while the \textit{q} of the resonance seems to 
provide immediately the ratio of the amplitudes for discrete and 
continuum excitation (and is often used as such \cite{tom}).  
However, as shown by Eq. (8), this simple 
correspondence does not exist.  To show the difference more clearly 
we rewrite Eq. (8) as 
\begin{equation}\label{9}
q = {\langle \phi | \mu | i \rangle 
\over \pi V^*_E \langle \psi_E|\mu | i\rangle_{\rm max}\cos\phi_i }
- \tan\phi_i ,
\end{equation}
showing that $q$ depends on both the ratio of the amplitudes to the 
discrete state and the unperturbed continuum and the 
phase $\phi_i$ between the initial state and the unperturbed continuum.  
In the absence of excitation to the discrete state
$q = -\tan\phi_i$ and can take any value.

In conclusion, we have observed symmetric peaks in a photoionization 
spectrum which appear to be due to excitation of only the discrete state. 
However, they are due to pure continuum excitation, in particular, to 
the phase shifted continuum $\Lambda_E$, since excitation to the 
unperturbed continuum $\psi_E$ vanishes. More generally, pure continuum 
excitation can lead to the entire range of Fano profiles so that an 
asymmetric lineshape does not necessarily imply interference between the
bound and continuum
excitation amplitudes. Consequently, the \textit{q} of an observed 
resonance does not, by itself, tell us the ratio of the excitation 
amplitudes to the discrete state and the unperturbed continuum. 
While we have here described the photoexcitation at a single resonance 
using an adaptation of Fano's configuration interaction theory which is 
applicable to long range potentials, we can also readily develop a quantum 
defect theory description extending from below the $5f$ limit to above 
the $5g$ limit.

It is a pleasure to acknowledge helpful conversations with W. Sandner, 
M. Fowler, R.R. Jones and H.J. Weber.  This work has been supported by the 
Deutsche Forschungsgemeinschaft and U.S. National Science Foundation, 
through grants PHY-9987948, DMR-9978074 (R.K.) and INT-9909827.

\end{document}